\documentclass[twocolumn,secnumarabic,amssymb, nobibnotes, aps, prl]{revtex4-1}

\usepackage{graphicx}
\usepackage{subfigure}
\usepackage{float}
\usepackage{color}
\usepackage{hyperref}
\begin{document}

\title{Single-gap superconductivity and dome of superfluid density in Nb-doped SrTiO$_3$}%

\author{Markus Thiemann$^1$, Manfred H. Beutel$^1$, Martin Dressel$^1$, Nicholas R. Lee-Hone$^2$,\\ David M. Broun$^{2,3}$, Evangelos Fillis-Tsirakis$^4$, Hans Boschker$^4$, Jochen Mannhart$^4$ and Marc Scheffler$^{1}$}%

\affiliation{$^{1}$1. Physikalisches Institut, Universit\"at Stuttgart, 70569 Stuttgart, Germany\\
$^2$Department of Physics, Simon Fraser University, Burnaby, BC, V5A 1S6, Canada\\
$^{3}$Canadian Institute for Advance Research, Toronto, Ontario, MG5 1Z8, Canada\\
$^{4}$Max Planck Institute for Solid State Research, 70569 Stuttgart, Germany
}
\date{\today}%
\begin{abstract}

SrTiO$_3$ exhibits a superconducting dome upon doping with Nb, with a maximum critical temperature \mbox{$T_\mathrm{c} \approx 0.4$~K}. Using microwave stripline resonators at frequencies from 2 to 23~GHz and temperatures down to 0.02~K, we probe the low-energy optical response of superconducting SrTiO$_3$ with charge carrier concentration from 0.3 to $2.2\times 10^{20}$~cm$^{-3}$, covering the majority of the superconducting dome. We find single-gap electrodynamics even though several electronic bands are superconducting. This is explained by a single energy gap $2\Delta$ due to gap homogenization over the Fermi surface consistent with the low level of defect scattering in Nb-doped SrTiO$_3$. Furthermore, we determine $T_\mathrm{c}$, $2\Delta$, and the superfluid density as a function of charge carrier concentration, and all three quantities exhibit the characteristic dome shape.
\end{abstract}

\maketitle

Amongst the numerous distinctive properties of SrTiO$_3$ \cite{mueller-burkard_prb_1979, Szot2006}, its superconducting state is of particular interest \cite{Schooley1964}. Stoichiometric SrTiO$_3$ is an insulating perovskite, but when charge carriers are introduced, e.g.\ by doping with Nb, SrTiO$_3$ becomes metallic with Fermi-liquid properties \cite{Spinelli2010, Lin2015,marel_prb_2011}. Such doped SrTiO$_3$ features several electronic bands at the Fermi level that can be filled consecutively, making SrTiO$_3$ a model system for multi-band physics \cite{marel_prb_2011,behnia_prl_2014}. This includes superconductivity in SrTiO$_3$ \cite{binnig_prl_1980, behnia_prl_2014}, which can reach a critical temperature $T_\mathrm{c}$ around 0.4~K, and is an important reference for the superconducting interface in SrTiO$_3$/LaAlO$_3$ heterostructures \cite{reyren-mannhart_science_2007}. Of particular interest is the evolution of superconductivity with doping: already a very small charge carrier density $n$ of $5.5 \times 10^{17}$~cm$^{-3}$ suffices to induce superconductivity \cite{Lin2013}. Only surpassed by superconducting Bi \cite{Prakash_Science2017}, this charge carrier density places SrTiO$_3$ in the non-adiabatic regime of superconductivity that is characterized by a Fermi velocity too low for conventional phononic coupling \cite{Behnia2017, Swartz_arxiv_2016}, thus leaving the mechanism for superconductivity in SrTiO$_3$ a matter of ongoing discussion \cite{Edge2016, Stucky2016}. Particularly relevant for our study is the evolution of $T_\mathrm{c}$ as a function of Nb doping. With increasing charge carrier density, $T_\mathrm{c}$ first increases, but for densities larger than $1 \times 10^{20}$~cm$^{-3}$ decreases again \cite{Lin2013, schooley-koonce_prl_1965, koonce-schooley_prl_1967}.
Such behavior, a dome-shaped evolution of $T_\mathrm{c}$ as function of a tuning parameter, is found in the phase diagrams of numerous superconducting material classes. These include cuprate, heavy-fermion, organic, pnictide, granular, and interface superconductors \cite{Jerome1991, MathurNature1998, Lefebvre_PRL2002, BrounNaturePhys2008, Caviglia2008, RichterNature2013, ShibauchiRevCon2014, Chen_PRL2015, PrachtPRB2016}, and the nature of such superconducting domes remains in the focus of scientific activity. One aspect is the relation between $T_\mathrm{c}$ and other energy scales that are relevant for the superconducting state, such as the superconducting energy gap $2\Delta$ and the superfluid stiffness $J$ that is proportional to the superfluid density $\rho_s$. In particular, for various superconducting systems that exhibit domes of $T_\mathrm{c}$ and $\rho_s$, the causal relationship between these two quantities remains a highly controversial issue \cite{PrachtPRB2016, Uemura_PRL1989, Emery_Nature1995, Homes_Nature2004, Zuev_PRL2005, Broun_PRL2007, Luan_PRL2011, Bozovic_Nature2016,Lee-Hone_PRB2017}. Here superconducting SrTiO$_3$ is an ideal model system to study the interplay between $T_\mathrm{c}$ and superfluid density: doping with Nb allows convenient control of electronic material properties including tuning through the superconducting dome. Furthermore, the role of disorder, which often complicates interpretation of composition-tuned superconducting domes, is well understood for SrTiO$_3$. Finally, because of the small energy gap, all relevant parameters such as $2\Delta$ and $\rho_s$ can be obtained from a single (optical) experiment on the same specimen, as we show in this work.

Optical spectroscopy at THz and infrared frequencies is an established route for the investigation of superconductors with $T_\mathrm{c}$ of a few K and higher \cite{Basov2005, PrachtIEEE2013}, but for superconductivity in SrTiO$_3$ one has to consider much lower relevant temperatures and frequencies. Consequently, we employ a microwave technique, namely superconducting stripline resonators suitable for the mK range \cite{SchefflerPSS2013, hafner_rsi_2014}. Since this spectroscopic technique probes the full complex electrodynamics of a superconductor, it directly indicates the superconducting gap $2\Delta$ and it quantifies the response of both the superfluid and the thermally excited quasiparticles. 
Combining our microwave investigation with normal-state transport measurements allows us to clearly elucidate the connections between pairing, phase stiffness, disorder, and multi-band electronic structure in Nb-doped SrTiO$_3$. 

\begin{figure}
\centering
\includegraphics[width=1\linewidth]{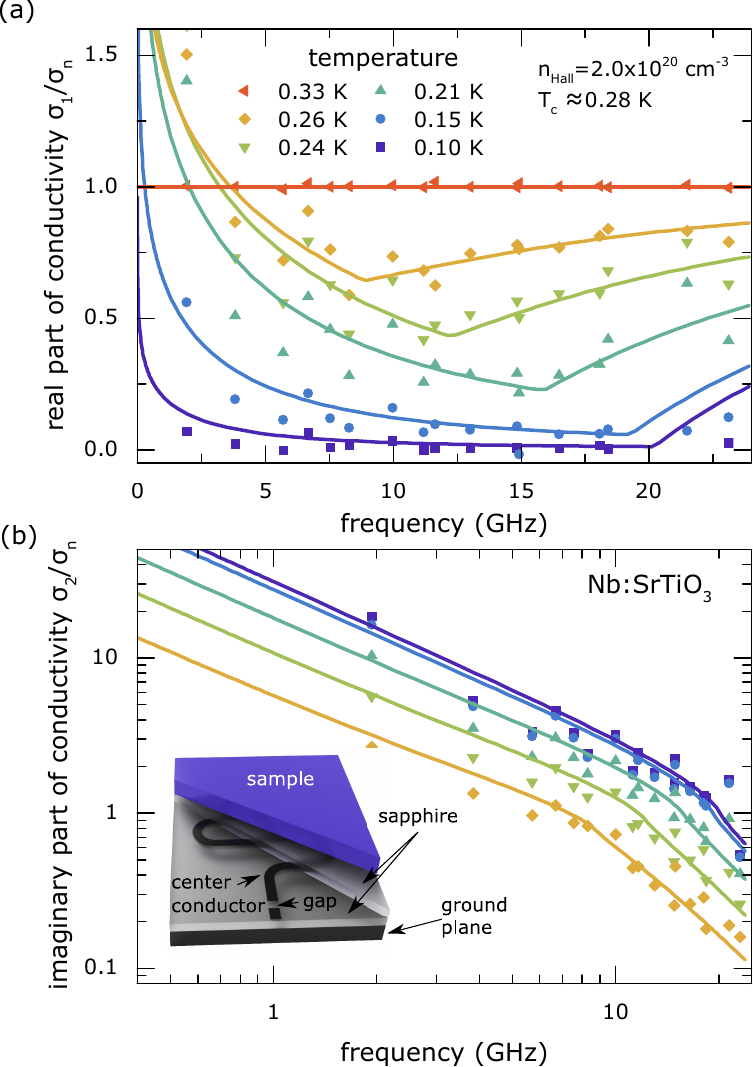}
\caption{Frequency dependence of the (a) real and (b) imaginary parts of the complex optical conductivity $\sigma=\sigma_1+\mathrm{i}\sigma_2$ in superconducting Nb-doped SrTiO$_3$ ($n_\mathrm{Hall}=2.0 \times 10^{20}~\mathrm{cm}^{-3}$, $T_\mathrm{c}\approx 0.28~\mathrm{K}$) plotted for different temperatures. Kinks in $\sigma(\omega)$ mark the spectroscopic gap 2$\Delta$. The solid lines represent  Mattis--Bardeen fits carried out simultaneously to $\sigma_1(\omega)$ and $\sigma_2(\omega)$ with the spectroscopic gap $2\Delta$ as the only free parameter. The inset shows a schematic drawing of the stripline resonator used to make the measurements. }
\label{fig:sigmavsf}
\end{figure}

In our stripline resonator configuration the Nb-doped SrTiO$_3$ samples act as a ground plane \cite{hafner_rsi_2014, SchefflerActaIMEKO2015}, as shown schematically in the inset of Fig.~\ref{fig:sigmavsf}(b). Using a cavity perturbation analysis, we obtain the surface impedance of the Nb-doped SrTiO$_3$ for a wide frequency range of \mbox{$\omega/2 \pi = 2$ to 23~GHz} and at temperatures down to $T = 0.02$~K, smoothly crossing from \mbox{$\hbar \omega \ll k_B T$ to $\hbar \omega \gg k_B T$}.  In all cases the microwave penetration depth is greater than 1~$\mu$m, allowing the measurements to probe deeply into the superconducting bulk. Details on the samples, microwave measurements and analysis are discussed in the supplemental material \cite{SupplementalMaterial}. 

The microwave surface impedance  gives direct access to the complex conductivity, $\sigma = \sigma_1 + \mathrm{i} \sigma_2$, which in turn encodes the electrodynamic response of both quasiparticles and superfluid. At sub-gap frequencies, the only absorption mechanism contributing to $\sigma_1$ comes from thermally excited quasiparticles, and is exponentially small at low temperatures.  Increasing the probing frequency further, pair-breaking sets in at \mbox{$\hbar\omega=2\Delta$}, leading to kinks in $\sigma_1(\omega)$ and $\sigma_2(\omega)$ at this frequency, as seen in Fig.~\ref{fig:sigmavsf}, which displays conductivity spectra of the sample with $n_\mathrm{Hall} = 2.0 \times 10^{20}$~cm$^{-3}$ for several temperatures.
Simultaneously fitting the Mattis--Bardeen equations \cite{mattis-bardeen_1958} to the real and imaginary parts of the conductivity spectra provides our first means of determining $2\Delta$, with results plotted as open symbols in Fig.~\ref{fig:keygraph2}(a). On a more fundamental level, the agreement of experimental spectra and Mattis--Bardeen fits suggests that the electrodynamic response of superconducting SrTiO$_3$ is governed by a single superconducting gap. We also point out that the pronounced downturn in $\sigma_2(\omega)$ above the gap frequency (See Fig.~\ref{fig:sigmavsf}b) is a subtle indication that disorder is relevant in this material \cite{SupplementalMaterial}.

\begin{figure}
\centering
\includegraphics[width=\linewidth]{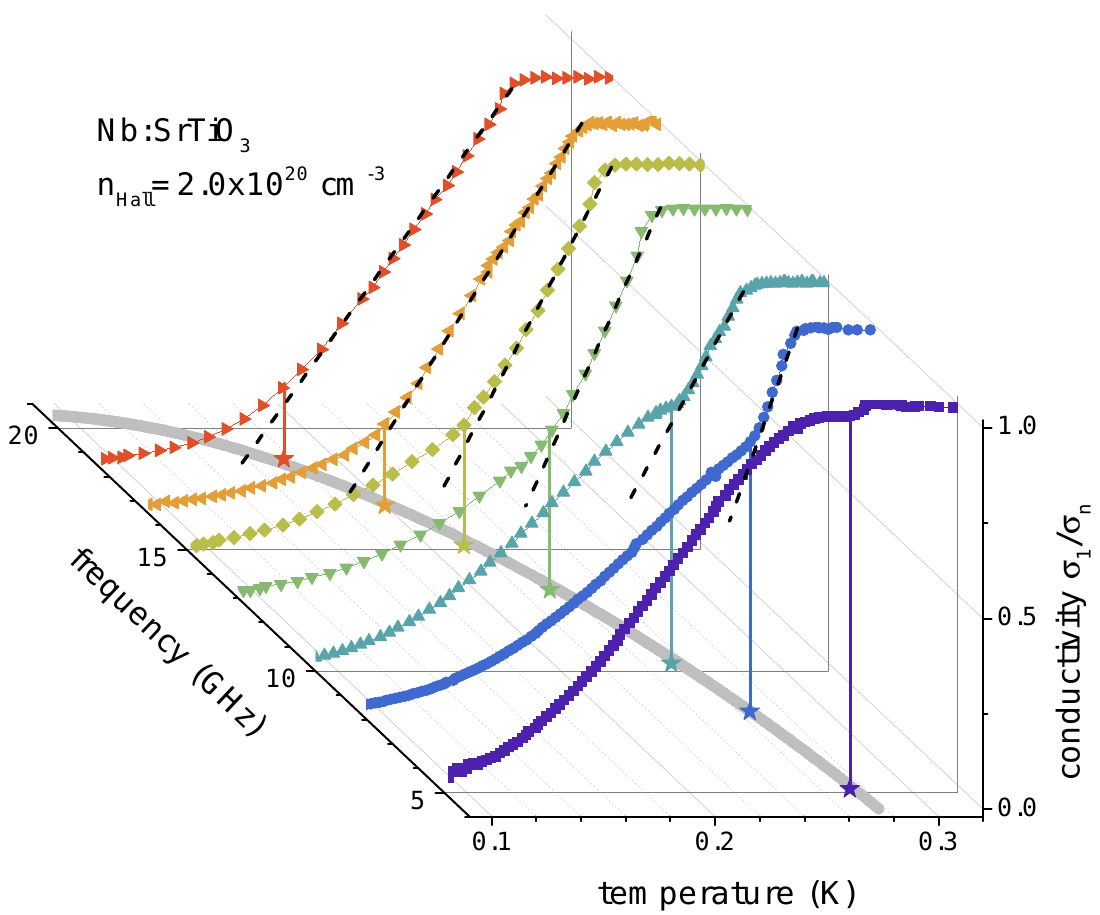}
\caption{Temperature dependence of $\sigma_1$, measured at various fixed frequencies for the sample shown in Fig.~\ref{fig:sigmavsf}. Kinks in $\sigma_1(T)$, marked by the lines projecting down to the frequency--temperature plane, indicate the temperature at which the microwave photon energy $\hbar \omega$ equals $2\Delta$. Dashed lines help to clarify the kinks. The grey line in the frequency--temperature plane shows the single-gap BCS prediction for $2\Delta(T)$.}
\label{fig:keygraph1}
\end{figure}

\begin{figure}
\centering
\includegraphics[width=\linewidth]{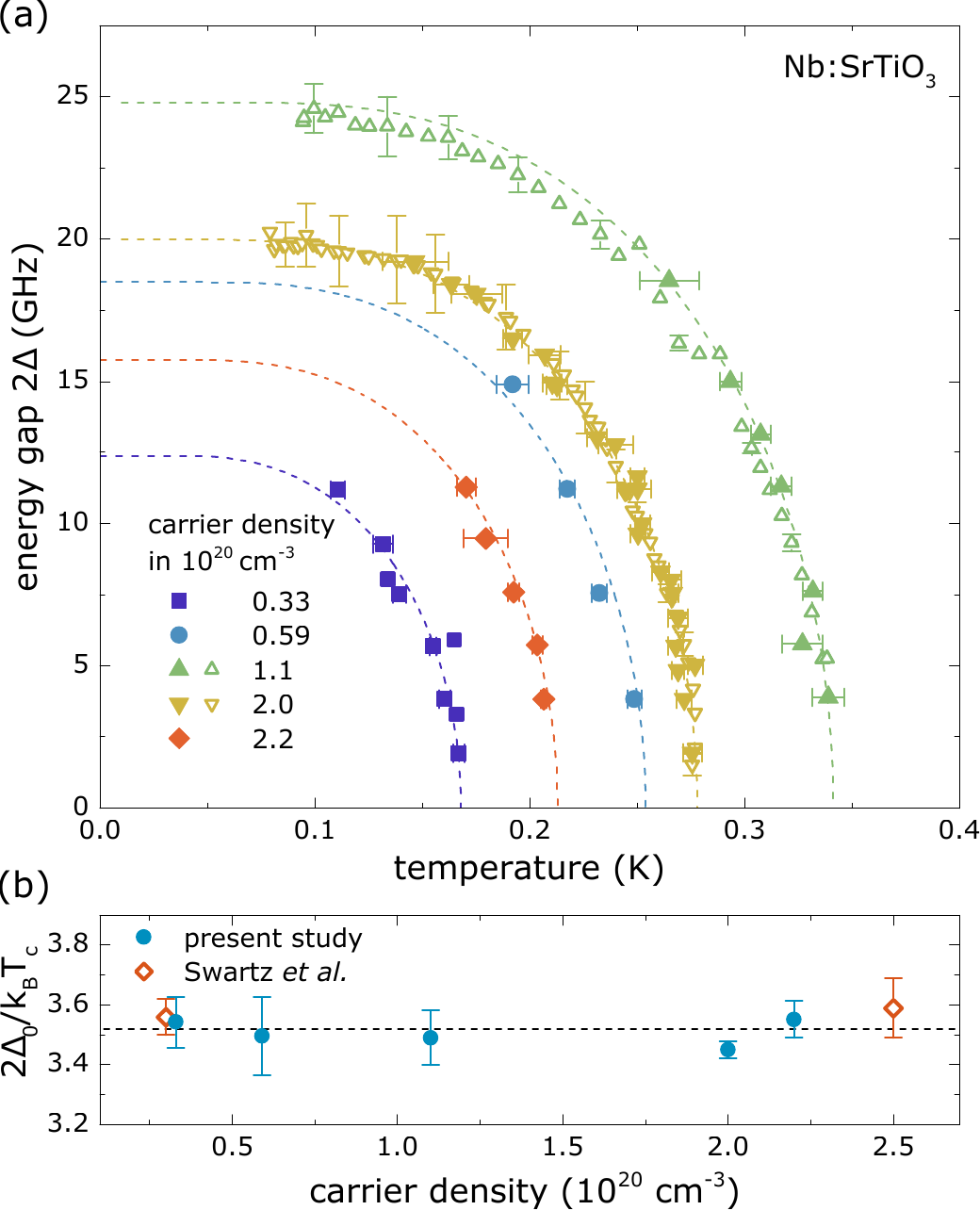}
\caption{(a) Temperature and doping dependence of the spectroscopic gap $2\Delta$: open symbols denote values from Mattis--Bardeen fits as shown in Fig.~\ref{fig:sigmavsf}; closed symbols are obtained from $\sigma_1(T)$ as shown in Fig.~\ref{fig:keygraph1}. Solid lines indicate fits to a single-gap BCS temperature dependence in which the zero-temperature gap, $2\Delta_0$, is the only adjustable parameter. (b) BCS gap ratio $2\Delta_0/k_BT_\mathrm{c}$ as a function of carrier density: closed symbols are from microwave spectroscopy; open symbols are from tunneling spectroscopy on thin films \cite{Swartz_arxiv_2016}; dashed line denotes the BCS value, $2\Delta_0/k_BT_\mathrm{c} = 3.528$.}
\label{fig:keygraph2}
\end{figure}

An independent way to extract $2\Delta(T)$ from our data, without relying on Mattis-Bardeen theory in detail, is by analyzing the temperature dependence of $\sigma_1$ for a set of frequencies, namely the resonator harmonics that we have available. For any given frequency $\omega$ that is smaller than the zero-temperature energy gap $2\Delta_0$, the temperature-dependent $\sigma_1(T)$ will indicate pronounced additional losses due to Cooper-pair breaking as the temperature is scanned through the point at which $2\Delta(T)$ falls below $\hbar \omega$. This is evident as kinks in $\sigma_1(T)$ in Fig.~\ref{fig:keygraph1}, which shows data for seven different resonator frequencies and $n_\mathrm{Hall}=2.0 \times 10^{20}$~cm$^{-3}$. The temperature dependence of the energy gap inferred by this method is plotted as the stars in the frequency-temperature plane of Fig.~\ref{fig:keygraph1} and as solid symbols in Fig.~\ref{fig:keygraph2}(a). 

The complete set of $2\Delta(T)$ is shown for all five samples in Fig.~\ref{fig:keygraph2}(a).
These data are well described by the temperature dependence of $2\Delta$ predicted for a single-gap BCS superconductor \cite{TinkhamBook}, where we treat the gap ratio $2\Delta_0/k_\mathrm{B}T_\mathrm{c}$ as an adjustable parameter.  Fitted gap ratios are plotted in Fig.~\ref{fig:keygraph2}(b) and are in close accord with the BCS prediction for weak-coupling superconductivity, $2\Delta_0/k_\mathrm{B}T_\mathrm{c} = 3.528$ \cite{TinkhamBook}. 
This value is also in line with recent tunneling results for doped SrTiO$_3$ thin films \cite{Swartz_arxiv_2016}. However, these findings are at odds with the early tunnelling data for doped SrTiO$_3$, which appeared to resolve multiple gaps \cite{binnig_prl_1980} (as was observed for other multiband superconductors \cite{IavaronePRL2002, RubyPRL2015}), but recently were interpreted in terms of surface superconductivity \cite{Eagles_JSNM2017}. Nevertheless this still leaves us with the question, why there is only one energy gap while there is strong evidence for multiple superconducting bands \cite{behnia_prb_2014}. 

\begin{figure*}[!ht]
\centering
\includegraphics[width=\linewidth]{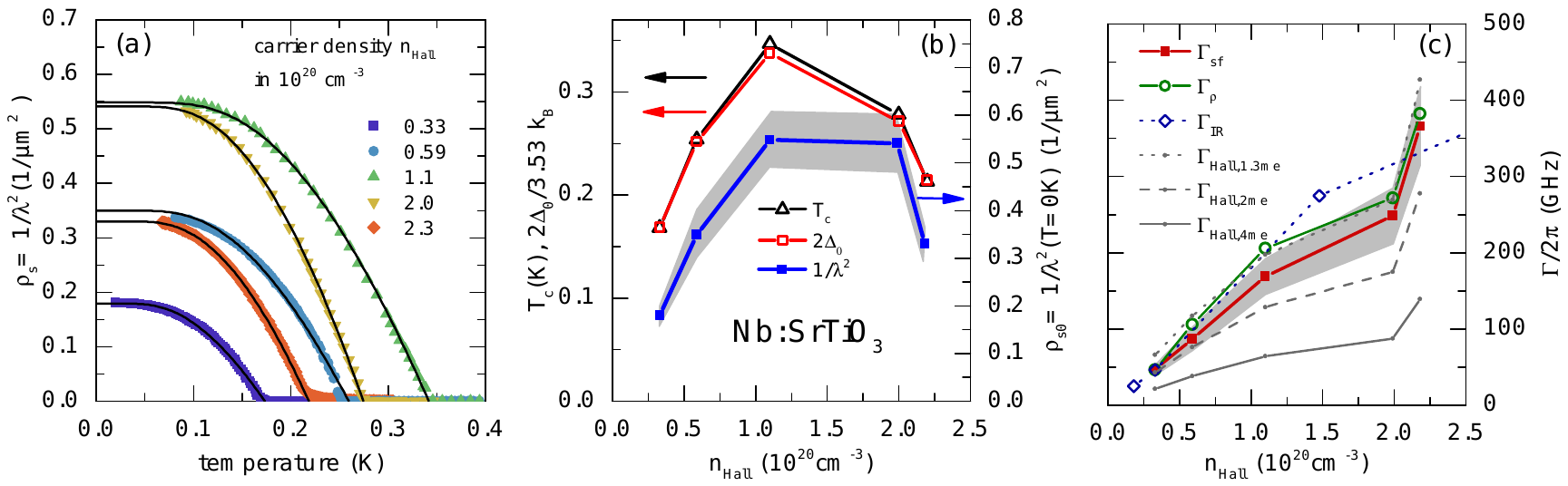}
\caption{(a) Temperature dependence of the absolute superfluid density $\rho_s$ for all five charge-carrier concentrations, measured at $\omega/2\pi\approx2$~GHz.  As described in the supplemental material \cite{SupplementalMaterial}, solid lines are fits to single-gap BCS superfluid density in the presence of disorder, with the absolute scale for the superfluid density taken from infrared measurements of the Drude weight \cite{vanMechelen2008}; scattering rate $\Gamma_\mathrm{sf}$ and $T_\mathrm{c}$ are the only two adjustable fit parameters.  (b) Domes of $T_\mathrm{c}$, $2\Delta_0$ and zero-temperature superfluid density $\rho_{s0}$ as a function of the charge carrier concentration determined from Hall effect measurements. The shaded area denotes the confidence band on  $\rho_{s0}$, including a 10\% scale uncertainty arising from the resistivity measurements used to calibrate the surface impedance. (c) The superfluid fit parameter $\Gamma_\mathrm{sf}$ is compared to independent determinations of scattering rate from infrared spectroscopy $\Gamma_\mathrm{IR}$ \cite{marel_prb_2011}, resistivity $\Gamma_\rho$, and Hall effect $\Gamma_\mathrm{Hall}$ assuming effective masses $m^*=1.3,2,4 m_\mathrm{e}$. The shaded area denotes the confidence band on $\Gamma_\mathrm{sf}$.}
\label{fig:sup_density}
\end{figure*}

A resolution of this puzzle emerges from Anderson's theorem \cite{Anderson1959}, which describes how conventional superconductivity is protected from nonmagnetic impurity scattering; disorder scattering does not suppress $s$-wave pairing, but instead homogenizes the energy gap over the Fermi surface.
In the case of a multiband superconductor with a scattering rate $\Gamma$ larger than the superconducting gap(s), this homogenization causes a single value for the spectroscopic gap throughout the complete Fermi surface even if the different bands had well separated, distinct values in the absence of scattering. (See supplemental material for more details \cite{SupplementalMaterial}.)
To support this concept for our actual case, we have to consider the scattering rates that are present in our samples. Electronic scattering in doped SrTiO$_3$ is typically quantified from the normal-state Hall mobility $\mu$ (see supplemental material \cite{SupplementalMaterial}), from which we extract a scattering rate via  $\Gamma_\mathrm{Hall} \equiv e/(\mu\, m^*)$ in a one-band interpretation and assuming several relevant values of the effective mass, \mbox{$m^* = 1.3, 2$ and $4~m_e$} based on quantum oscillation measurements \cite{behnia_prl_2014}, with the heavier one expected to dominate in our dopant range. Furthermore, using the normal-state resistivity $\rho_\mathrm{DC}$ of our samples and the plasma frequency $\omega_p$ estimated from optical studies of Nb-doped SrTiO$_3$ \cite{vanMechelen2008}, we can determine a transport scattering rate $\Gamma_{\rho}=\epsilon_0\omega_p^2\rho_\mathrm{DC}$ 
which in Fig.~\ref{fig:sup_density}(c) is plotted together with comparable values $\Gamma_\mathrm{IR}$ for the samples of Refs.~\cite{marel_prb_2011, vanMechelen2008} and $\Gamma_\mathrm{Hall}$. For all our samples, these scattering rates are larger than the respective superconducting energy gaps, thus validating our explanation of single-gap superconductivity caused by scattering in a multi-band system.
Furthermore, the roughly linear increase of $\Gamma$ with $n_\mathrm{Hall}$ indicates that the dominant scattering mechanism is impurity scattering due to the Nb dopants.
 
It is important to place these ideas in the context of field-dependent measurements of thermal conductivity \cite{behnia_prb_2014} and surface impedance (see supplemental material \cite{SupplementalMaterial}), which indicate clearly that multiple bands contribute to superconductivity.  These field-dependent probes reveal two distinct scales: the upper critical field, $B_{c2} = \Phi_0/2 \pi \xi^2$, the point at which vortex spacing reaches the coherence length $\xi$ and superconductivity is globally destroyed; and a lower field scale, $B^\ast$, indicating an additional, longer superconducting coherence length.  In a multiband system, band-specific coherence lengths $\xi_{i} \equiv \hbar v_{\mathrm{F},i}/\pi \Delta_i$ (index $i$ for different bands) \cite{TinkhamBook} naturally arise from band-to-band variation in either the energy gap, $\Delta_i$, or Fermi velocity, $v_{\mathrm{F},i}$.  In Nb-doped SrTiO$_3$, quantum oscillation studies \cite{behnia_prl_2014} indicate enough variation in $v_{\mathrm{F},i}$ that we expect multiple field scales to emerge even in the presence of the homogeneous energy gap implied by the spectroscopic measurements.

Further insights into the structure of the superconducting gap come from the temperature dependence of superfluid density. In previous studies \cite{Kogan_PRB2009}, multi-band gap structure in $\rho_s(T)$ exhibited either as pronounced, mid-range upwards curvature, e.g., in V$_3$Si \cite{Nefyodov_epl_2005}, or as the presence of small activation energies, e.g., in MgB$_2$ \cite{Fletcher_prl_2005} and FeSe \cite{Teknowijoyo_PRB2016, Meng_njp_2016}.  
In our measurements, superfluid density is obtained from the low frequency limit ($\omega/2\pi\approx 2$~GHz) of the out-of-phase conductivity, $\rho_\mathrm{s}(T) \equiv \lambda^{-2}= \omega\mu_0\sigma_2(\omega,T)$, where $\lambda$ is the London penetration depth \cite{TinkhamBook}.  
We note that for low $T_\mathrm{c}$ superconductors such as Nb-doped SrTiO$_3$, electrodynamic measurements are an excellent means for determining absolute superfluid density, as the superconducting penetration depth is measured relative to the high-frequency  skin-depth in the nearby normal state, providing a reliable reference point that can be calibrated in terms of dc resistivity measured on the same sample \cite{pippard}.
The temperature dependences of the absolute superfluid density of all five samples are shown in Fig.~\ref{fig:sup_density}(a), along with fits to $\rho_s(T)$ from single-gap, weak-coupling BCS in the presence of nonmagnetic disorder (see supplemental material \cite{SupplementalMaterial}).  
These fits provide additional experimental evidence for single-gap behavior, and the values $\Gamma_\mathrm{sf}$ for the scattering rate that we obtain from the fits are consistent with the other estimates shown in Fig.~\ref{fig:sup_density}(c) and thus confirm our interpretation of single-gap superconductivity in terms of Anderson's theorem.

Furthermore, the zero-temperature limits $\rho_{s0}$ of the fits for $\rho_s$ allow us to study the evolution of the superfluid density as a function of normal-state charge carrier density. In Fig.~\ref{fig:sup_density}(b) we plot $\rho_{s0}$ versus $n_\mathrm{Hall}$, and we find a dome shape rather similar to the domes of $T_\mathrm{c}$ and $2\Delta_0$. A dome of superfluid density in Nb-doped SrTiO$_3$ has also recently been reported in Ref.~\onlinecite{Collignon2017}. As additional context, Fig.~S5 shows that Nb-doped SrTiO$_3$ follows the Homes' law scaling between $\rho_s$ and $\sigma_\mathrm{DC}T_\mathrm{c}$. 

Following the Ferrell-Glover-Tinkham sum rule, $\rho_{s0}$ is proportional to the spectral weight in $\sigma_1(\omega)$ that is lost in the superconducting state compared to the normal state \cite{TinkhamBook}. For all our samples the scattering rate is larger than $2\Delta$, placing them in the dirty limit, and thus the transferred spectral weight is limited by $2\Delta$ (or $T_\mathrm{c}$). Consequently the $T_\mathrm{c}$ dome in SrTiO$_3$ is not governed by $\rho_{s0}$.
This result is particularly interesting in the context of the ongoing discussion concerning the superconducting domes in other material classes, most notably the cuprates.
These domes of superconductivity are also often accompanied by $\rho_{s0}$ behavior that qualitatively tracks the rise and fall of $T_\mathrm{c}$. However, the causal relationship between $\rho_s$ and $T_\mathrm{c}$ remains a point of major controversy, in particular whether superfluid density places bounds on $T_\mathrm{c}$ or, instead, $T_\mathrm{c}$ controls the spectral weight available to form the superfluid \cite{Bozovic_Nature2016,Emery_Nature1995,Kogan_PRB2013,Lee-Hone_PRB2017}.  
Our findings demonstrate that for SrTiO$_3$ the latter explanation holds and that the electronic scattering due to disorder has to be considered for a full understanding of the superfluid density. 
This scattering rate in superconducting SrTiO$_3$, which is very small for a metallic system but unavoidable due to the required charge carrier doping, is also a crucial ingredient for our observation that Nb-doped SrTiO$_3$ is a single-gap, multi-band superconductor.

We thank G. Untereiner for resonator and sample preparation and K. Behnia and H. Y. Hwang for helpful discussions.
Financial support by Carl-Zeiss-Stiftung (M.T.), DFG, and DAAD is thankfully acknowledged. D.M.B. and N.R.L.-H. gratefully acknowledge financial support from the Natural Science and Engineering Research Council of Canada and the Canadian Institute for Advanced Research. M.T. and M.H.B. contributed equally to this work.

\end{document}